# Unexpected water-hydroxide ion structure and diffusion behavior in low hydration media

I. Zadok[1], D. R. Dekel[1,2*], and S. Srebnik[1*]

Affiliations:

[1] Department of Chemical Engineering, Technion – Israel Institute of Technology, Haifa 3200003, Israel

[2] The Nancy & Stephan Grand Technion Energy Program (GTEP), Technion – Israel Institute of Technology, Haifa 3200003, Israel

*Correspondence to: simchas@technion.ac.il, dario@technion.ac.il

## Abstract

Hydroxide ion transport and structure in aqueous media is fundamental to many chemical and biological processes. Research on hydroxide behavior has primarily focused on a single fully solvated hydroxide, either as an isolated cluster or in the bulk. This work presents the first computational study to consider a medium of low hydration levels where the hydroxide ion is microsolvated. Under such conditions, hydroxide ions are shown to be predominantly present as unique water-bridged double-hydroxide charged clusters, distinct from previously reported structures under hydrated conditions. Although layered double hydroxides were reported in the crystalline state, this is the first time to be seen in the disordered liquid state. These newly observed double-hydroxide structures presumably disrupt the hydrogen bonded network required for structural diffusion of hydroxide ions through water. These ion complexes have a higher ionic strength which may explain the unexpected diffusion behavior in comparison to the single hydroxide-water complex.

# Main

Solvation of protic species, specifically hydroxyl(hydroxide) and oxonium(hydronium) moieties, plays a major role in many fields and application, from conformation and activity of proteins[1] in biological systems to stability of electrochemical systems such as anion exchange membrane fuel cells[2]. Solvation of hydronium cations in aqueous medium has been extensively studied and much is known about these systems[3]. However, our understanding of solvation of hydroxide anions in aqueous medium is in its infancy[4–6], having primarily focused on the structure and transport of a single ion in aqueous medium.

Transport of hydroxide ions in aqueous medium is typically divided into two distinct mechanisms – vehicular and structural diffusion. Vehicular diffusion is the movement of the center of charge together with the center of mass together. Structural diffusion involves movement of the center of charge separately from the center of mass, usually realized by proton ($H^+$) transfer from one distinct molecule the another, also known as "Grotthuss diffusion", which alters the center of charge whilst keeping the center of mass mostly unchanged. The structure and transport of hydroxide in aqueous medium are interconnected. To a first approximation, the diffusion constant is inversely proportional to radius of the diffusing particle and is determined by the dominant solvated structure[7]. The solvated structure is specified by the arrangement of water molecules, distance, and orientations, around the central hydroxide oxygen. These water molecules can be divided into solvation layers according to their distance from the oxygen. Thus, factors that affect the solvated structure are expected to influence both vehicular and structural diffusion mechanisms of the hydroxide ion.

Zatsepina[8] first evaluated the structure of hydroxide in alkali metal hydroxide solution using NMR and Raman spectra and concluded that hydroxide exists mainly as $[HO^-\cdots H^+\cdots OH^-]^-$ complex with one water molecule. Buchner at el.[9] used dielectric relaxation measurements of NaOH solutions to estimate the coordination number of OH⁻ at infinite dilution as 5.5. Botti at el.[10] corrected this value using neutron scattering and molecular modelling to 3.9 neighboring oxygens per hydroxide, in accord with the classical Lewis acid evaluation of OH⁻ having three accepting hydrogen bonds (HBs) using the three lone atom pairs and one donating bond[5]. However, density functional theory (DFT) - molecular dynamics (MD) simulations carried out by Tuckerman and coworkers[11] on a single fully solvated hydroxide ion revealed that these accepted HBs are non-localized. The HB electrons form a torus shapes orbital around the hydroxide oxygen. Thus, OH⁻ can accept four HBs and donate a single weaker HB, leading to hyper-coordination of OH⁻. Chen at el.[12] simulated hydroxide-water solutions with different concentrations of Na⁺/K⁺ counter ions. Their findings slightly differ and point to three hydroxide-water structures, OH⁻ $(H_2O)_x$, where $x$=3, 4, and 5, whose distribution is a function of the counter ion. In an effort to reconcile experimental measurements of the activation energy for hydroxide mobility, Agmon[7] argued the dominant structures to be 3-fold coordinated structure OH⁻$(H_2O)_3$ and a dimer anion [HO⁻--H⁺--OH⁻]⁻, so that the high coordination numbers measured are in fact a result of averaging the first solvation shell of the two oxygens around the dimer anion.

Due to the added effect of structural diffusion, hydroxide (and hydronium) show large mobility in bulk. While for hydronium a simple mechanism of chain proton transfer was found to explain both qualitatively and quantitively this behavior (i.e., the Grotthuss mechanism)[6], applying the exact mechanism to hydroxide ion results in inaccuracies both in structure and diffusion coefficient[13]. Depending on the functional used, two different mechanisms of structural diffusion using proton transfer were observed[13]. The first is the Grotthuss mechanism, where a proton transfers along a chain of OH⁻$(H_2O)_3$ complexes which resulted in a diffusion coefficient an order of magnitude higher than the experimental value. The second mechanism involves a water molecule leaving the hyper-coordinated structure, the transfer of a proton from one of the water molecules in the first solvation layer, and recreation of the hyper-coordination around the new center of charge. The latter mechanism predicted a diffusion coefficient closer to the experimental value. Agmon[7] on the other hand, explained hydroxide mobility using a unified model similar to hydronium, where the dimer anion discussed above is the facilitating structure for proton transfer. However, little experimental data is available to corroborate the various predictions of these first principle models.

The focus of these studies was mostly on a single hydroxide ion in bulk water. Hydroxide ion structure and transport under microsolvated conditions, where there is not enough water to fully solvate the hydroxide ion, has not yet been reported, to the be best of our knowledge. While very little attention has been paid to hydroxide ions in low hydration levels, several practical systems work in this particular environment. Microsolvation can occur naturally in non-aqueous systems where water is present as an impurity, or in hydrated environment where water is constantly consumed by chemical or electrochemical reaction[2]. Performance and stability characteristics of these systems are significantly affected by the very low hydration medium. For instance, it has been recently shown that the high hydroxide reactivity observed in experiments on quaternary ammonium (QA) cations results from the lack of full solvation of the hydroxide ions[14]. In spite of the increasing importance of these systems, to date hydroxide ion structure and transport in low hydration levels have not been studied.

In this work, we focus on hydroxide and water transport in in low hydration media, relating the local structures of hydration and hydrogen bonding with the macro view of hydrophilic clusters and the diffusion coefficient. We investigate the behavior of hydroxide ions in a model QA system under such conditions using atomistic MD simulation. Focusing on hydroxide-water interactions, we investigate the hydroxide ion structure at hydration levels ranging from full water solvation to microsolvation and anhydrous conditions. Implications of our findings towards hydroxide transfer dynamics of the solvated complexes are discussed. To the best of our knowledge, this is the first time the behavior of microsolvated hydroxide is investigated.

**Results and Discussion**

When solvated, hydroxide ions diffuse via a combination of structural and vehicular diffusion [5,6]. However, the diffusion mechanism of hydroxide ions under microsolvated conditions is still unknown. Such conditions have been shown to be critical in determining the properties of anionic conducting polymers[2]. Diffusion coefficients for water and hydroxide in the ternary system consisting of (**1**) water, (**2**) hydroxide, (**3**) and benzyl triethylammonium (BTEA$^+$) cation, for various degrees of hydration (defined as water:hydroxide ratio, or $\lambda$) is plotted in Figure 1. Hydroxide ion diffusivity for the entire range of hydration levels is shown for the first time, to the best of our knowledge.

As expected, diffusivity rapidly decrease as hydration decreases[15]. A decrease of around two orders of magnitude is seen between values of diffusivity in bulk and in ultra-low hydration conditions. Remarkably, the decrease in diffusivity is not uniform along the hydration scale, and below a certain value of $\lambda$, diffusion deviates from the expected behavior. Two distinct diffusivity zones are observed: diffusion at a low hydration level ($4 < \lambda < 15$), and diffusion at an ultra-low hydration level ($\lambda < 4$). Prediction of the free volume diffusion model (described in Methods) are shown in solid lines. The model successfully captures the variability in the diffusion, as well as predicts the knot value of free volume separating the two distinct diffusivity zones (ca. $\lambda = 3\text{-}4$). Model fit to equation (4) (see Methods) and corresponding knot values, activation energy, the critical free volumes, estimated errors, and $R^2$ values are tabulated in Supporting Information Table S1.

Figure 1 also shows the calculated diffusivity of hydroxide ion in bulk water ($\lambda \rightarrow \infty$, dashed lines) and the respective experimental values reported in the literature (open symbols).

Very good agreement is observed for the diffusivity of the water. The difference between the calculated diffusivity and reported experimental diffusivity of the hydroxide ion may be due to the absence of structural diffusing in the classical model. However, as we will argue below, vehicular diffusion is likely to dominate at low hydration.

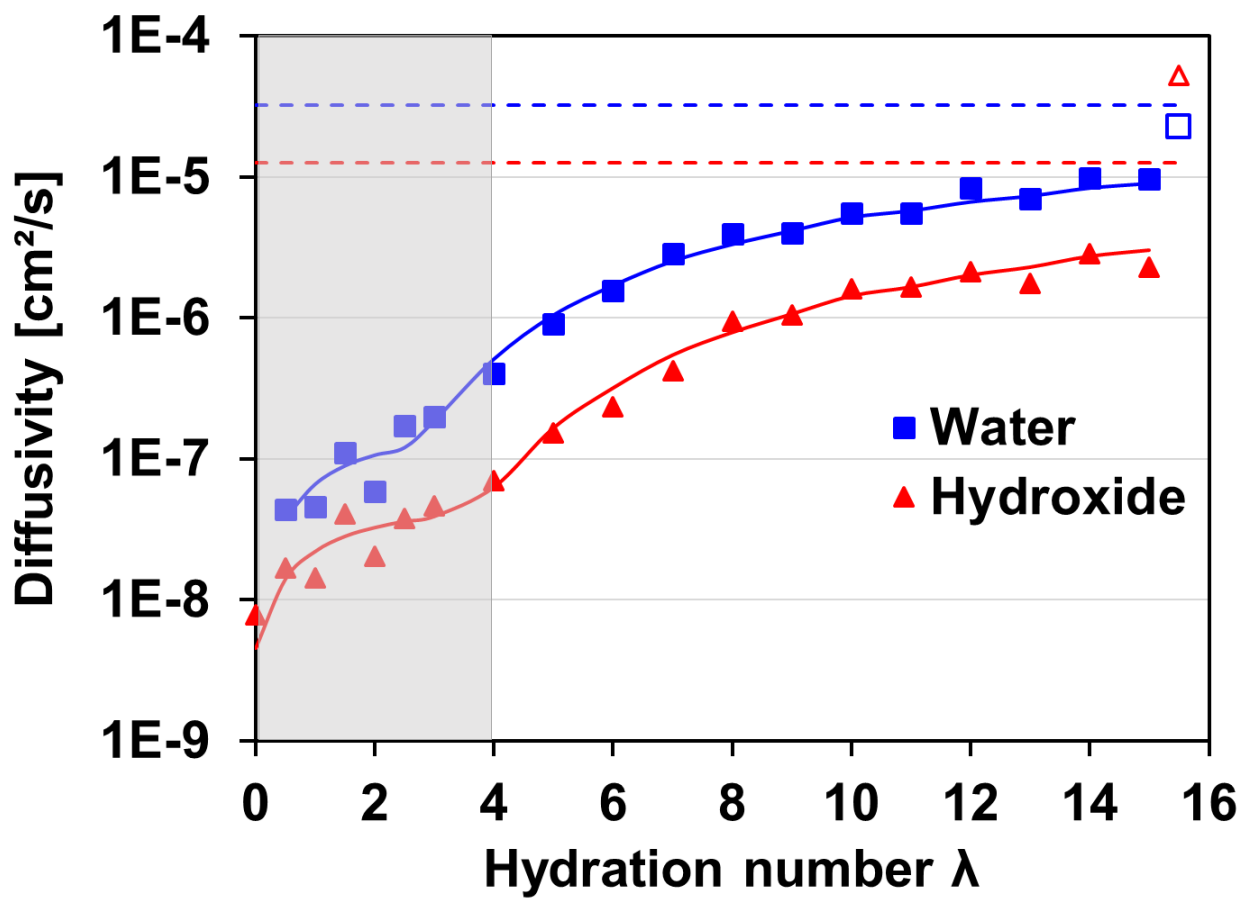

Figure 1. Simulated diffusion coefficients of water (squares) and hydroxide ion (triangles) for various hydration levels. Filled symbols correspond to simulations carried out at 313K. Solid lines correspond to the free volume diffusion model given by equation (4). Dashed lines correspond to the diffusion coefficients calculated at bulk aqueous conditions. Open symbols correspond to experimental bulk data measured at 298K[16,17].

The different diffusion regimes at low and ultra-low hydration levels shown in Figure 1 are believed to be an outcome of the different structure of water-hydroxide complexes. Figure 2 shows the average number of HBs per hydroxide ion and water, at different hydration levels. For hydroxide, at ultra-low hydration levels a steep rise in the number of HBs is observed with increasing λ. In addition, the number of HBs per hydroxide ion saturates at ~4.5 HBs around λ = 6-8, in agreement with several computational predictions for hydrated (hyper-coordinated) hydroxide[13,18]. For water, on the other hand, HBs begin to evolve only around λ = 2. However, even at relatively high hydration (λ = 15), the number of HBs is still far from the bulk value of 1.9 and reported simulation results 2.08 HBs[19].

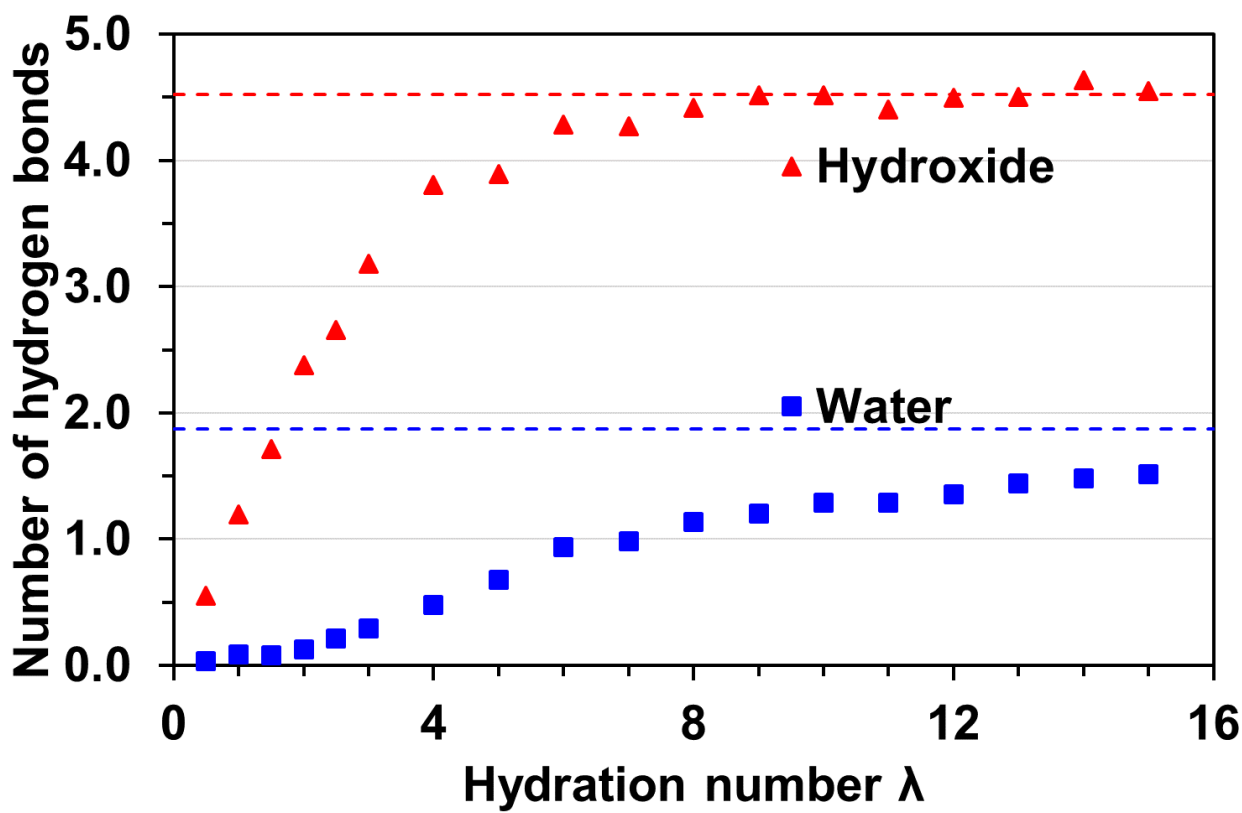

Figure 2. Average number of hydrogen bonds per OH⁻ and per water molecule as a function of λ. Dashed lines are values simulated at bulk conditions.

Marcus[20] summarized the effects of different ions on the HB structure of water. Simulation of tetramethyl ammonium in dilute (λ = 750) water solution[21] found that while the water oxygen-oxygen radial distribution function does not change with the addition of QA, the HB angle (∠(O•••O-H)) distribution is narrowed, producing a stronger HB network. This is analogous to the findings of Imberti and coworkers[22], who discovered that counterions (K⁺, Na⁺ or Li⁺) have a remarkable effect on the structure of the formed hydroxide water clusters and subsequently on the water HBs. Marcus classified ions as structure-making or structure-breaking ions[23] according to an empirical estimate of Gibbs energy of HB formation. Hydroxide is classified as structure-making while short chain alkylammonium ions are structure-breaking. However, the Marcus scale does not consider cooperative or competing effects, nor microsolvation effects. Our findings reveal a large structure-breaking effect even at hydration levels where the hydroxide ion is fully solvated ($\lambda > 6$). These results, combined with the large difference in hydration enthalpies (-35.52 *kJ/mol*[24] versus -472.6 *kJ/mol*[25] for water and hydroxide, respectively) confirm that hydroxide competes favorably against the formation of a water HB network.

Figure 3 shows snapshots of hydrophilic clusters (in the timeframe containing the largest cluster formed during the simulation) for various hydration levels and the corresponding HB network within the largest cluster. The internal structure of the clusters is seen to be composed of hydrated hydroxide interconnected by a network of water molecules. At ultra-low hydration ($\lambda = $ 1-1.5) many small unconnected spherical and elongated clusters exist. As hydration increases ($\lambda = $ 2-3), a large branched cluster begins to dominate, although most of the hydrophilic region is still primarily composed of isolated clusters. Clear percolation of a single hydrophilic cluster is seen at $\lambda = 4$, enabling molecules to diffuse freely across the simulation box boundaries. This value corresponds to the transitional hydration level between the two regimes of diffusion observed in Figure 1. For $\lambda < 4$, lack of percolation of the hydrophilic clusters requires diffusion of the clusters for transport to occur, stressing the importance of vehicular diffusion in this regime. Diffusion at ultra-low hydration levels is therefore governed by cooperative motion of water-hydroxide clusters.

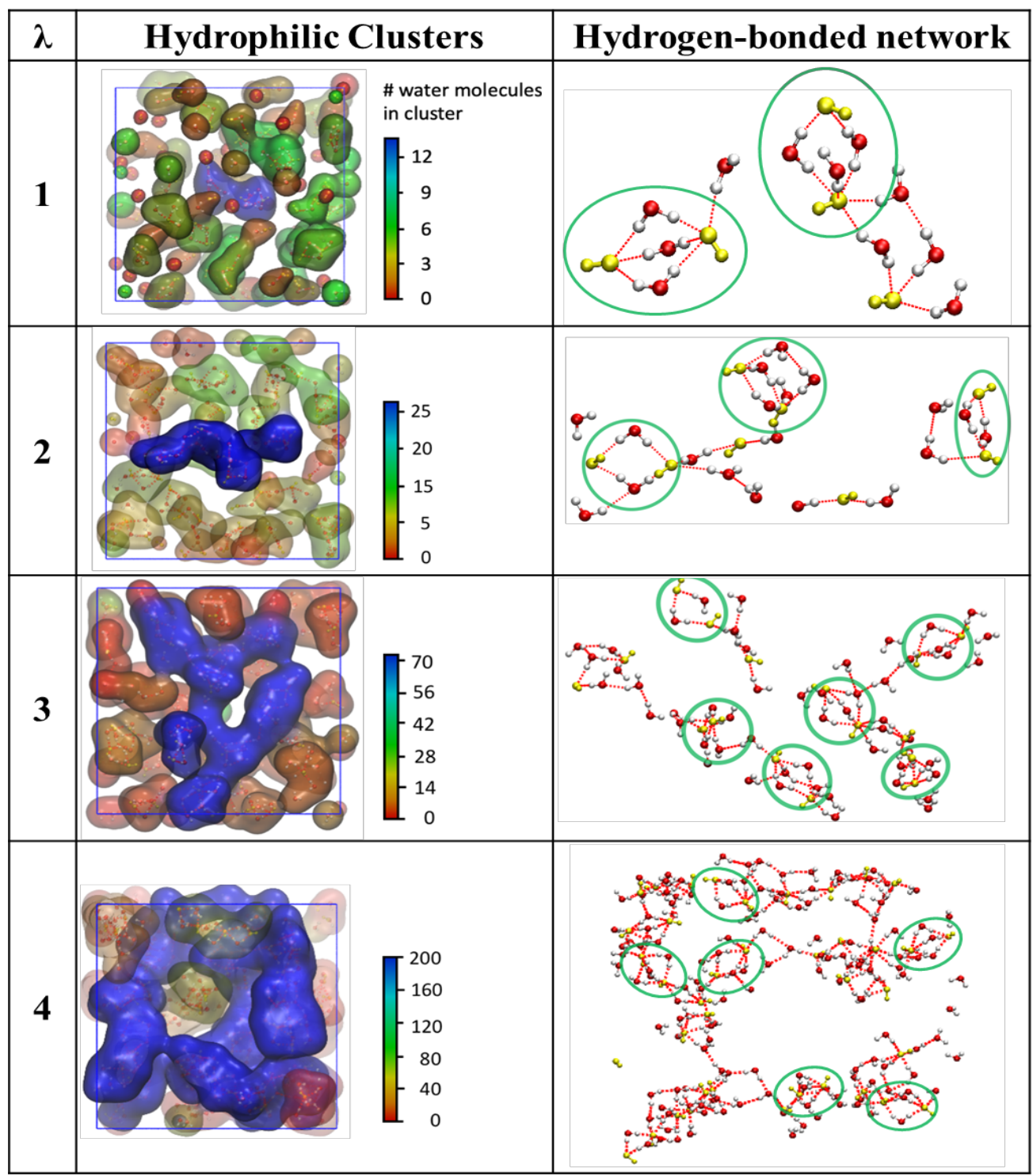

*Figure 3. Size and shape of hydrophilic clusters for different hydration levels and corresponding hydrogen bonded network of the largest hydrophilic clusters. Clusters are colored according to size: small clusters in red, intermediate clusters in green, and large clusters in blue. Depth cuing is used for 3D representation. Hydroxide is displayed in yellow, water oxygens in red, and water hydrogens in white. Hydrogen bonds are shown as dotted red lines. Formation of peculiar water bridged double-hydroxide structures, $O_4H_6^{2-}$ and $O_5H_8^{2-}$, are marked by green circles.*

Careful observations of the HB network of the clusters at the ultra-low hydration regime reveals two recurring divalent structures composed of two hydroxides bridged by two or three water molecules – a (quasi) planner rhombus ($O_4H_6^{2-}$) and a trigonal dipyramid ($O_5H_8^{2-}$). These structures correspond to the spherical hydrophilic clusters seen in Figure 3 and act as nodes which are interconnected by elongated regions that correspond to a network of hydrogen bonded water molecules. DFT geometric optimization was used to confirm the stability of these structures at 0K. The optimized rhombus and dipyramid double hydroxide structures are shown in Figure 4a-4b. Figure 4c shows the optimized classical structure of hydrated hydroxide $O_5H_9^-$ for comparison. Surprisingly, the rhombus and dipyramid structures are of higher ionic strength (estimation included in supplementary information) than the conventional single hydroxide hydrated structure ($O_5H_9^-$). We note that the rhombus structure was previously observed experimentally in the crystalline state of tetraethylammonium hydroxide pentahydrate[26] at a temperature of 213K, while the trigonal dipyramid was observed in hybrid quantum mechanics/MD simulations in bulk water without counter ions[27].

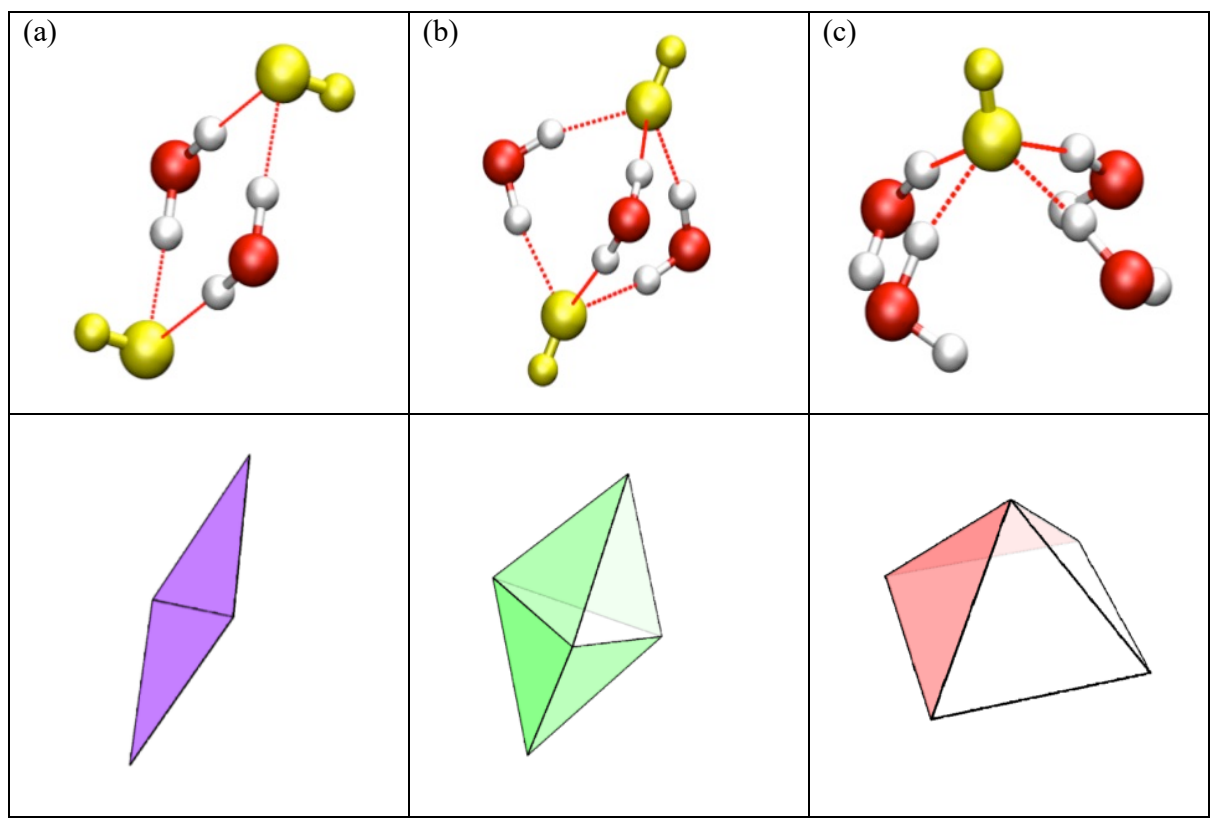

*Figure 4. Structures of geometrically optimized water-hydroxide ion complex: (a) $O_4H_6^{2-}$ rhombus double hydroxide, (b) $O_5H_8^{2-}$ trigonal dipyramid double-hydroxide, and (c) hyper-coordinated $O_5H_9^-$ single hydroxide coordinated by 4 water molecules (as predicted by Tuckerman et al.[13]). Schematics representing the geometric shapes were constructed using Wolfram Mathematica 11.3. Hydroxides are displayed in yellow, water oxygens in red, and water hydrogens in white.*

Notably, most of the current research on hydroxide transport in aqueous medium focuses on the hydration of single hydroxides or dilute solutions[5,13,28], hence the water bridged double-hydroxide structures, $O_4H_6^{2-}$ and $O_5H_8^{2-}$, have not been reported in this context since these will be entropically less favorable. The presence of a strong positive QA supporting charge in the concentrated ionic system under consideration compensates for the increase in coulombic repulsion in these double-hydroxide structures. Interestingly, in a study of hydroxide transfer in QA-functionalized membrane at intermediate hydration, Voth et al.[29] concluded that most of the hydroxide transport takes place in the first solvation layer of QA, and that a considerable component of the transport mechanism is derived from vehicular diffusion as opposed to Grotthuss diffusion. The predominance of these new structures may shed further light on this phenomenon.

Local structures were further analyzed using the radial distribution function $g(r)$ and average number of neighbors $CN(r)$ as a function of distance for pairs of atoms of hydroxide and water for various $\lambda$ (Figure 5). Coordination layers appear as spherical shells from valley-to-valley in the $g(r)$ plot, where the number of atoms at each shell is given by the differential value of $CN(r)$ from previous layers. It is important to note that these plots do not imply hydrogen-bonded atoms, which require a higher order correlation. Nonetheless, the structure of water around hydroxide is illuminated using the correlation $g(r)_{Oh-Hw}$ of the hydroxide oxygen and water hydrogen (Figure 5b). While the most probable distance (highest peak) of finding a water molecule around the hydroxide remains roughly the same for all hydration levels, the population of water molecules within this first shell ($r = 1.5$-$2.4$Å) changes with hydration, indicated by the corresponding plateau in $CN(r)$. We note that while $CN(r)$ for the first hydration shell (first plateau in Figure 5a) under bulk conditions is 6, only 4 of these are actual HBs (see Figure 2).

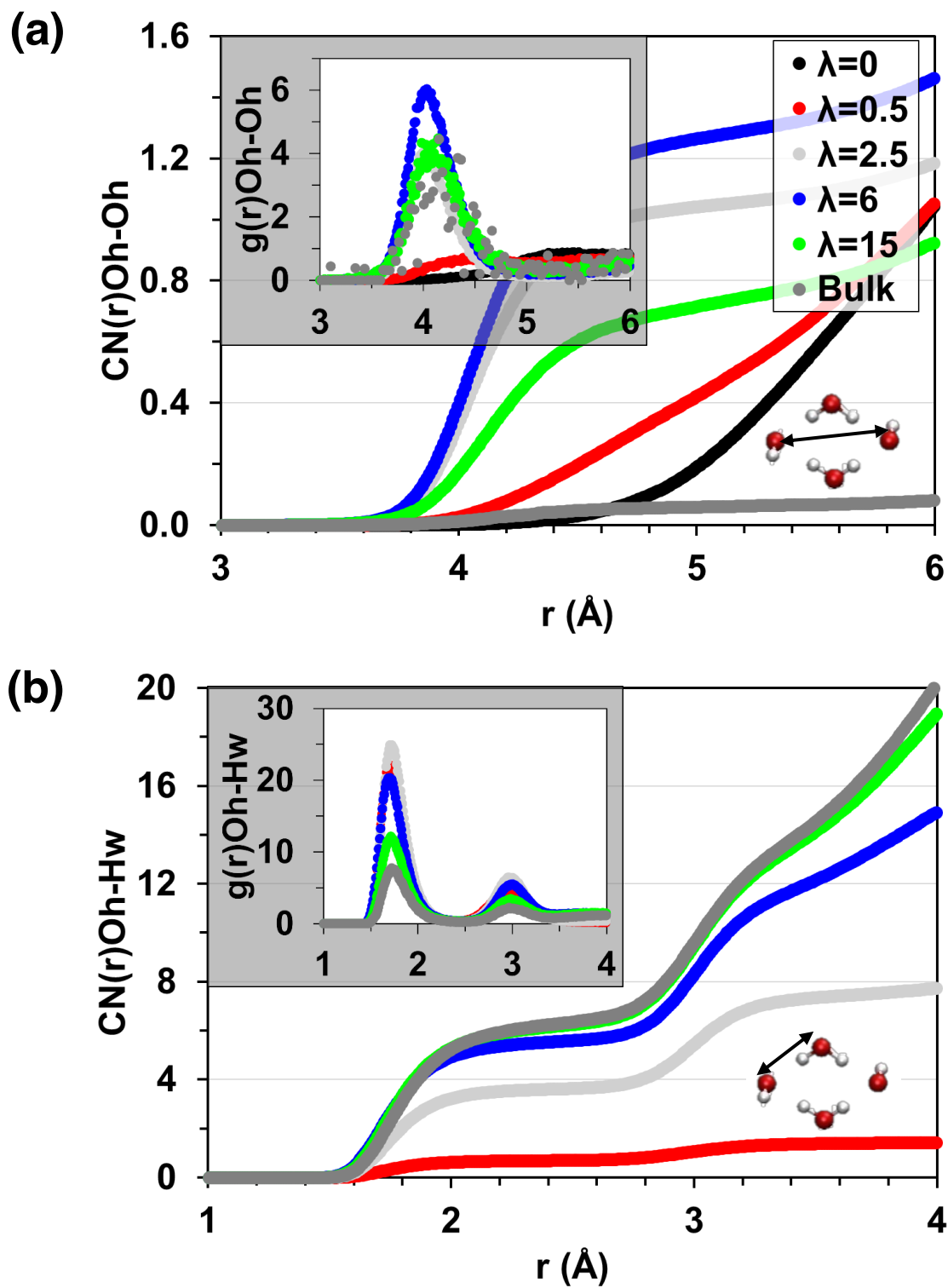

*Figure 5. Running coordination number CN(r) for (a) Oh-Oh (hydroxide oxygen and hydroxide oxygen) and (b) Oh-Hw (hydroxide oxygen and water hydrogen). Corresponding radial distribution functions g(r) are shown in the insert on the top left; illustration of the measured distance is shown in the insert on the bottom right.*

The radial distribution of hydroxide oxygens (Figure 5a) allows us to quantify the extent of the water-bridged double hydroxide structures. It is seen that for ultra-low hydration numbers there is little correlation between the hydroxides, but as hydration increases a peak around 4Å emerges with a value of $CN \approx 1.2$. This distance is associated with the water-bridged hydroxide structures shown in Figure 4a-4b. As hydration is further increased beyond $\lambda = 6$, the frequency of appearance of these structures diminishes ($CN$ decreases). At $\lambda = 15$, $CN$ drops to 0.7, and drops further to 0.1 in bulk. That is, hydroxides in bulk water are more likely to be isolated and uncoordinated from each other, as *a priori* assumed by previous computational studies which focused on single hydroxide in various degrees of hydration [7,11,13,28]. The occupancy of these water-bridged ionic states relative to the occupancy of non-correlated states is clearly a function of hydration, and is probably influenced by the type of cation[10,22]. Indeed, the population of hydroxide-water complexes was shown to be influenced by the cation type for alkali hydroxides[12]. In the system under consideration, bridged states are not observed at ultra-low hydration levels due to the strong correlation with the ammonium cations. As hydration increases, the hydroxide is shielded by its hydration layer and the occupation of the bridged structures increases. As the ion concentration gets further diluted, the bridged hydroxide-water structures are entropically less likely to occur.

The occurrence of these newly observed hydroxide ion structures is likely to dramatically influence the diffusive behavior of hydroxide at low hydration, so that its dynamic properties should be viewed as an ensemble average of the two forms. Although structural diffusion was not studied in this work, the dominance of the bridged structure at ultra-low hydration suggests that the mechanisms proposed by Agmon[7] and Tuckerman[13] would be less prevalent under these conditions, either due to the added enthalpy required for separating the two hydroxides, or due to the highly coordinated mechanism needed to achieve the required configuration for structural diffusion.

**Conclusions**

Atomistic molecular dynamics simulations were used to study the structure and transport of hydroxide ions in water under a wide range of hydration conditions in the presence of quaternary ammonium cations. Low and ultra-low hydration conditions are shown to be dominated by a largely unexplored water-bridged double hydroxide ion structure for which dynamic properties are unknown. In this work, hydroxide ion diffusivity in the ultra-low hydration regime was explored for the first time, revealing the existence of two distinctive diffusive regimes whose transport properties are likely to be strongly influenced by the population of these water-bridged double hydroxide structures. A revised free volume diffusion model describes the simulated data. A newly found transition in the diffusivity of hydroxide and water was revealed at hydration levels around $\lambda = 2.5$-$4$, indicating different diffusive behavior between the ultra-low and low hydration regimes, with important implications towards hydroxide ion behavior under microsolvated conditions. Several parallel processes associated with hydration were identified. The ultra-low hydration regime was characterized by isolated small hydrophilic clusters whose lack of connectivity limits diffusion. This hydration regime is dominated by the water-bridged double hydroxide structures, which were found to be of higher ionic strength than the hydrated single hydroxide structure. At higher hydration levels ($\lambda > 4$), diffusivity transitions to a different regime as hydrophilic regions percolate and a hydrogen bonded network is formed. The water-bridged double-hydroxide ion structures diminish in their concentration with further hydration but are still prominent.

In summery, in low and ultra-low hydration media, we found an unusual hydroxide ion structure and diffusion behavior. Our results show that in this medium, the microsolvated hydroxides are structured as unique water-bridged double-hydroxide ions. We expect that vehicular diffusion should dominate in this microsolvated ion regime, and will be higher than that expected for a simple single hydrated ion. Our findings bring new light to the transport mechanisms of hydroxide ions in systems with low water content.

## Methods

*Molecular Dynamics Simulation*

We studied the behavior of BTEA$^+$ (**3**) hydroxide (**2**), shown on the right, under different hydrated conditions. The ions were modeled with the OPLS force field[30] and water was modeled with TIP5P[31] (TIP5P was selected for its good accuracy in modeling diffusivity[32]). MD simulations were carried out using LAMMPS[33]. Cation:water ratios of 1:0-1:15 (corresponding to hydration levels of $0 \leq \lambda \leq 15$) were considered, as well as dilute conditions ($\lambda \approx 100$-$200$). Initial configurations of the ion-water system were constructed using the enhanced Monte Carlo (EMC) code[34]. The system was equilibrated at three temperatures (313, 333, and 353K) and at 1 atm under isothermal-isobaric ensemble using the Nosé-Hoover algorithm[35,36] for the thermostat and barostat, with the recommended time constants of 100 and 1000 fs, respectively. Cutoff for the pair interaction potentials was taken as 9.5Å. PPPM long-range correction[37] for Coulombic interactions was used with 10$^{-3}$ relative error. The system was equilibrated for 1ns using 1fs integration time-step. Analysis was carried out under the canonical ensemble for trajectories recorded over the course of 2 ns.

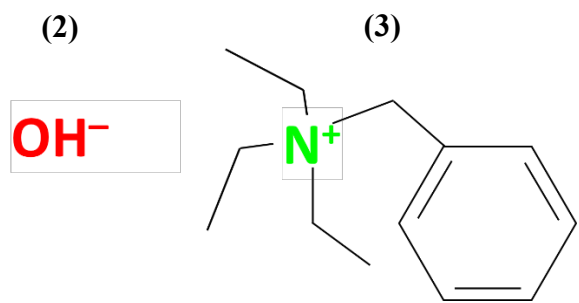

*Diffusion Model*

Two approaches are often used to describe the functional form of diffusion within this kind of systems. The thermodynamic approach (which utilizes the Onsager transport coefficients , i.e. Janssen[38], used for proton conducting polymeric systems) and the kinetic approach based on the concept of free volume diffusion[39], adapted for a two-component system (polymer and solvent) by Vrentas and Duda[40]. The kinetic approach was recently expanded for multicomponent systems by Ohashi[41], using a framework based on collision theory. We base our calculations on the kinetic approach due to its simplicity, direct relation to molecular properties, and direct functional dependence on external factors, such as free-volume.

Diffusivity of the water, hydroxide and ammonium ions were evaluated using the mean square displacement (MSD) and a time lag method[42] using VMD's *diffusion_coefficient* plugin[43], according to:

(1) $$MSD(\tau,\delta) = \frac{1}{N_\tau}\sum_{j=1}^{N_\tau}\left[\frac{1}{N_p}\sum_{i=1}^{N_p}\left|\overrightarrow{r_i(j\delta+\tau)} - \overrightarrow{r_i(j\delta)}\right|^2\right]$$

where $\tau$ is the time, $\delta$ is the time lag, $N_p$ is the number of particles, $N_\tau$ is the number of time lags, and $r_i$ is the vector from the origin.

The diffusion coefficient $D$ was then calculated from equation (2) as the average of ten largest lag times. A 95% confidence interval was calculated, and typical values were smaller than 1% of the average diffusion coefficient value.

(2) $$D(\delta) = \frac{1}{6}\lim_{\tau\to\infty}\frac{MSD(\tau,\delta)}{d\tau}$$

Diffusion coefficients were fitted to the free volume diffusion model (Macedo and Litovich[39]) for various liquid types including hydrogen bonded liquids, given by:

(3) $$D = D_0 \exp\left[-\frac{E}{RT}\right]\exp\left[-\frac{B}{\phi}\right]$$

where $D_0$ is the pre-exponential factor, $E$ is an activation energy, $R$ is the universal gas constant, $T$ is the temperature, $B$ is a specific critical free volume which also accounts for overlap of penetrate molecules, and $\varphi$ is the average free volume fraction.

Multiple linear regression was used to fit equation (2) divided by the reference diffusion coefficient at 25°C[16] on a log-log plot. The mobile phase volume ratio was used in place of the free specific volume, which in our system is the combined volume of water and hydroxide (molecular volumes were estimated using Voronoi tessellation using LAMMPS *compute voronoi/atom based on VORO++ code*[44]). Multiple linear regression for a single curve showed a parabolic shape for the residuals, indicating the actual data is better described using a piecewise linear regression model[45]. A "dummy" variable was added to the regression curve using a switching variable for the reciprocal hydrophilic volume. The knot value was used to maximize $R^2$ as a good measure of fit. The complete model used is given by:

(4)
$$\ln\left[\frac{D_i}{D_{ref}}\right] = \ln\left[\frac{D_0}{D_{ref}}\right] + \frac{E}{R}\left[-\frac{1}{T_i}\right] + B\left[-\frac{1}{\phi_{i1}}\right] + B^*\left[-\frac{1}{\phi_{i1}} + \frac{1}{\phi^*}\right]\phi_{i2} + \varepsilon_i$$

$$\phi_{i2} = \begin{cases} 0, & \phi_{i1} \leq \phi^* \\ 1, & else \end{cases}$$

where $D_{ref}$ is the diffusion coefficient (self or tracer) in pure water at 298K, $\phi^*$ is the knot value binding the two curves together, $B^*$ is the addition to the specific free volume above the knot value, the index $i$ runs over the piecewise regressions, $\phi_{i2}$ is a switching variable, and $\varepsilon_i$ are independent errors with zero mean. The fitted coefficients, regression statistics, residual plots, and normal probability plots are presented in the supplementary information.

The average number of HBs per hydroxide and per water was calculated using geometric method in VMD (*measure hbonds* command). HBs were defined according to distance and angle constraints of $r(O\cdots O)<3.5Å$ and $\angle(O\cdots O-H)\leq 30°$[19,46], respectively. The internal structure of the system was investigated using pair distribution functions[47] for the oxygen, hydrogen and nitrogen atoms. The distribution function is a measure of local density and its integral in the running coordination number. Pair and HB statistics were collected for 1000 time frames and averaged, a 95% confidence interval was calculated and typical values are lower than 1% of the average value.

*Density functional geometric optimization*

Small, micro-solvated hydroxide-water clusters were geometrically optimized using DFT in CP2K[48] using BLYP functional[49] and TZVP-GTH basis set[50]. Plane wave cutoff was set at 600 Ry, energy tolerance was set at $1e^{-11}$, and SCF tolerance at $1e^{-7}$.

## References

1. Steinbach, P. J. & Brooks, B. R. Protein hydration elucidated by molecular dynamics simulation. *Proc. Natl. Acad. Sci. U. S. A.* **90,** 9135–9 (1993).

2. Dekel, D. R. *et al.* Effect of Water on the Stability of Quaternary Ammonium Groups for Anion Exchange Membrane Fuel Cell Applications. *Chem. Mater.* **29,** 4425–4431 (2017).

3. Marx, D., Tuckerman, M. E., Hutter, J. & Parrinello, M. The nature of the hydrated excess proton in water. *Nature* **397,** 601–604 (1999).

4. Chen, M. *et al.* Hydroxide diffuses slower than hydronium in water because its solvated structure inhibits correlated proton transfer. *Nat. Chem.* **10,** 413–419 (2018).

5. Marx, D., Chandra, A. & Tuckerman, M. E. Aqueous Basic Solutions: Hydroxide Solvation, Structural Diffusion, and Comparison to the Hydrated Proton. *Chem. Rev.* **110,** 2174–2216 (2010).

6. Agmon, N. *et al.* Protons and Hydroxide Ions in Aqueous Systems. *Chem. Rev.* **116,** 7642–7672 (2016).

7. Agmon, N. Mechanism of hydroxide mobility. *Chem. Phys. Lett.* **319,** 247–252 (2000).

8. Zatsepina, G. N. State of the hydroxide ion in water and aqueous solutions. *J. Struct. Chem.* **12,** 894–898 (1972).

9. Buchner, R., Hefter, G., May, P. M. & Sipos, P. Dielectric Relaxation of Dilute Aqueous NaOH, NaAl(OH) 4 , and NaB(OH) 4. *J. Phys. Chem. B* **103,** 11186–11190 (1999).

10. Botti, A., Bruni, F., Imberti, S., Ricci, M. A. & Soper, A. K. Ions in water: The microscopic structure of concentrated NaOH solutions. *J. Chem. Phys.* **120,** 10154–10162 (2004).

11. Tuckerman, M. E., Marx, D. & Parrinello, M. The nature and transport mechanism of hydrated hydroxide ions in aqueous solution. *Nature* **417,** 925–929 (2002).

12. Chen, B., Ivanov, I., Park, J. M., Parrinello, M. & Klein, M. L. Solvation Structure and Mobility Mechanism of OH - : A Car−Parrinello Molecular Dynamics Investigation of Alkaline Solutions. *J. Phys. Chem. B* **106,** 12006–12016 (2002).

13. Tuckerman, M. E., Chandra, A. & Marx, D. Structure and Dynamics of OH - (aq). *Acc. Chem. Res.* **39,** 151–158 (2006).

14. Pusara, S., Srebnik, S. & Dekel, D. R. Molecular Simulation of Quaternary Ammonium Solutions at Low Hydration Levels. *J. Phys. Chem. C* **122,** 11204–11213 (2018).

15. Marino, M. G., Melchior, J. P., Wohlfarth, A. & Kreuer, K. D. Hydroxide, halide and water transport in a model anion exchange membrane. *J. Memb. Sci.* **464,** 61–71 (2014).

16. Easteal, A. J. & Woolf, L. A. Diffusion of Protonic Species in Aqueous NaOH as a Function of Concentration, Temperature, and Pressure, and Diffusion of Water in Aqueous NaF, NaBr, NaCIO, NaBF, NaNO, NaNO, and NaBrO, at 298 K and 0.1 MPa. *J . Phys. Chem* **90,** 2441–2445 (1986).

17. Sarode, H. N. *et al.* Insights into the Transport of Aqueous Quaternary Ammonium Cations: A Combined Experimental and Computational Study. *J. Phys. Chem. B* **118,** 1363–1372 (2014).

18. Zhu, Z. & Tuckerman, M. E. Ab Initio Molecular Dynamics Investigation of the

Concentration Dependence of Charged Defect Transport in Basic Solutions via Calculation of the Infrared Spectrum †. *J. Phys. Chem. B* **106,** 8009–8018 (2002).

19. Zielkiewicz, J. Structural properties of water: Comparison of the SPC, SPCE, TIP4P, and TIP5P models of water. *J. Chem. Phys. J. Chem. Phys. J. Chem. Phys. J. Chem. Phys. J. Chem. Phys.* **123,** 104501–234505 (2005).

20. Marcus, Y. Effect of Ions on the Structure of Water: Structure Making and Breaking. *Chem. Rev.* **109,** 1346–1370 (2009).

21. Madan, B. & Sharp, K. Changes in water structure induced by a hydrophobic solute probed by simulation of the water hydrogen bond angle and radial distribution functions. *Biophys. Chem.* **78,** 33–41 (1999).

22. Imberti, S. *et al.* Ions in water: The microscopic structure of concentrated hydroxide solutions. *J. Chem. Phys.* **122,** 194509 (2005).

23. Marcus, Y. ViscosityB-coefficients, structural entropies and heat capacities, and the effects of ions on the structure of water. *J. Solution Chem.* **23,** 831–848 (1994).

24. Mejías, J. A. & Lago, S. Calculation of the absolute hydration enthalpy and free energy of H+ and OH−. *J. Chem. Phys.* **113,** 7306–7316 (2000).

25. Tissandier, M. D. *et al.* The Proton's Absolute Aqueous Enthalpy and Gibbs Free Energy of Solvation from Cluster-Ion Solvation Data. *J. Phys. Chem. A* **102,** 7787–7794 (1998).

26. Wiebcke, M. & Felsche, J. NEt4 OH·5H 2 O containing hydroxide–water layers. *Acta Crystallogr. Sect. C Cryst. Struct. Commun.* **56,** 1050–1052 (2000).

27. Ghosh, M. K., Choi, T. H. & Choi, C. H. Like-charge ion pairs of hydronium and hydroxide in aqueous solution? *Phys. Chem. Chem. Phys.* **17,** 16233–16237 (2015).

28. Tuckerman, M., Laasonen, K., Sprik, M. & Parrinello, M. Ab Initio Molecular Dynamics Simulation of the Solvation and Transport of H3O+ and OH- Ions in Water. *J. Phys.*

*Chem.* **99,** 5749–5752 (1995).

29. Chen, C., Tse, Y.-L. S., Lindberg, G. E., Knight, C. & Voth, G. A. Hydroxide Solvation and Transport in Anion Exchange Membranes. *J. Am. Chem. Soc.* **138,** 991–1000 (2016).

30. Jorgensen, W. L., Maxwell, D. S. & Tirado-Rives, J. Development and testing of the OPLS all-atom force field on conformational energetics and properties of organic liquids. *J. Am. Chem. Soc.* **118,** 11225–11236 (1996).

31. Mahoney, M. W. & Jorgensen, W. L. Quantum, intramolecular flexibility, and polarizability effects on the reproduction of the density anomaly of liquid water by simple potential functions. *J. Chem. Phys.* **115,** 10758–10768 (2001).

32. Jorgensen, W. L. & Tirado-Rives, J. Potential energy functions for atomic-level simulations of water and organic and biomolecular systems. *Proc. Natl. Acad. Sci. U. S. A.* **102,** 6665–70 (2005).

33. Plimpton, S. Fast parallel algorithms for short-range molecular dynamics. *J. Comput. Phys.* **117,** 1–19 (1995).

34. Veld, P. J. in 't. EMC: Enhanced Monte Carlo. (2016). at <http://montecarlo.sourceforge.net/emc/Welcome.html>

35. Nosé, S. A unified formulation of the constant temperature molecular dynamics methods. *J. Chem. Phys.* **81,** 511–519 (1984).

36. Shinoda, W., Shiga, M. & Mikami, M. Rapid estimation of elastic constants by molecular dynamics simulation under constant stress. *Phys. Rev. B* **69,** 134103 (2004).

37. Hockney, R. & Eastwood, J. *Computer Simulation Using Particles*. (Taylor & Francis, 1988). doi:10.1201/9781439822050

38. Janssen, G. J. M. A Phenomenological Model of Water Transport in a Proton Exchange Membrane Fuel Cell. (2001). doi:10.1149/1.1415031

39. Macedo, P. B. & Litovitz, T. A. On the Relative Roles of Free Volume and Activation Energy in the Viscosity of Liquids. *J. Chem. Phys.* **42,** 245–256 (1965).

40. Vrentas, J. S. & Duda, J. L. Diffusion in polymer—solvent systems. I. Reexamination of the free-volume theory. *J. Polym. Sci. Polym. Phys. Ed.* **15,** 403–416 (1977).

41. Ohashi, H. & Yamaguchi, T. General Diffusion Model for Polymeric Systems Based on Microscopic Molecular Collisions and Random Walk Movement. *Ind. Eng. Chem. Res.* **52,** 9940–9945 (2013).

42. Pranami, G. & Lamm, M. H. Estimating Error in Diffusion Coefficients Derived from Molecular Dynamics Simulations. *J. Chem. Theory Comput.* **11,** 4586–4592 (2015).

43. Giorgino, T. Computing diffusion coefficients in macromolecular simulations: the Diffusion Coefficient Tool for VMD. (2015). at <https://github.com/tonigi/vmd_diffusion_coefficient/>

44. Rycroft, C. H. VORO++: A three-dimensional Voronoi cell library in C++. *Chaos An Interdiscip. J. Nonlinear Sci.* **19,** 041111 (2009).

45. Montgomery, D. C., Peck, E. A. & Vining, G. G. *Introduction to Linear Regression Analysis*. (Wiley, 2012).

46. van der Spoel, D., van Maaren, P. J., Larsson, P. & Tîmneanu, N. Thermodynamics of Hydrogen Bonding in Hydrophilic and Hydrophobic Media. *J. Phys. Chem. B* **110,** 4393–4398 (2006).

47. Allen, M. P. & Tildesley, D. J. *Computer Simulation of Liquids*. (Oxford University Press, 1989).

48. Hutter, J., Iannuzzi, M., Schiffmann, F. & VandeVondele, J. cp2k: atomistic simulations of condensed matter systems. *Wiley Interdiscip. Rev. Comput. Mol. Sci.* **4,** 15–25 (2014).

49. Becke, A. D. Density-functional exchange-energy approximation with correct asymptotic

behavior. *Phys. Rev. A* **38,** 3098–3100 (1988).

50. Goedecker, S., Teter, M. & Hutter, J. Separable dual-space Gaussian pseudopotentials. *Phys. Rev. B* **54,** 1703–1710 (1996).

## Acknowledgments

This work was partially funded by the Nancy & Stephan Grand Technion Energy Program (GTEP); by the European Union's Horizon 2020 research and innovation program [grant No. 721065]; by the Ministry of Science, Technology & Space of Israel through the Israel-Germany Batteries Collaboration Call 2017 [German grant No. 2675], and through grant No. 3-12948; by the Israel Science Foundation (ISF) [grant No. 1481/17]; by the Israel Innovation Authority through the KAMIN program [grant No. 60503]; and by the Ministry of National Infrastructure, Energy and Water Resources of Israel [grant No. 3-13671]. The authors would also like to acknowledge the financial support Planning & Budgeting Committee / ISRAEL Council for Higher Education (CHE) and Fuel Choice Initiative (Prime Minister Office of ISRAEL), within the framework of Israel National Research Center for Electrochemical Propulsion (INREP).

## Author information

DRD and SS generated the main initial ideas for the project; IZ carried out all the simulations and calculations and wrote most of the initial manuscript; DR and SS guided and edited the manuscript. IZ, DRD and SS discussed the analyses, conclusions, and contributed ideas and interpretation.